\documentclass[aps,showpacs,pra,amssymb, amsmath, twocolumn]{revtex4}
\usepackage{graphicx}
\usepackage{amssymb}
\usepackage{amsmath}
\usepackage{bm}
\begin{document}
\title{Robust two-level system control by a detuned and chirped laser pulse}
\author{Hanlae Jo,$^1$ Han-gyeol Lee,$^1$ St\'{e}phane Gu\'{e}rin,$^2$ and Jaewook Ahn$^1$}
\email{jwahn@kaist.ac.kr}
\address{$^1$Department of Physics, KAIST, Daejeon 305-701, Korea}
\address{$^2$Laboratoire Interdisciplinaire Carnot de Bourgogne, UMR CNRS 6303, Universit/'{e} Bourgogne Franche-Comt\'{e},
F-21078 Dijon Cedex, France}
\begin{abstract}
We propose and demonstrate a robust control scheme by ultrafast nonadiabatic chirped laser pulse, designed for targeting coherent superpositions of two-level systems.  Robustness against power fluctuation is proved by our numerical study and a proof-of-principle experiment performed with femtosecond laser interaction on cold atoms. They exhibit for the final driven dynamics a cusp on the Bloch sphere, corresponding to a zero curvature of fidelity. This solution is particularly simple and thus applicable to a wide range of potential applications.
\end{abstract}

\maketitle


Quantum information technologies are expected to play an important role in a near future~\cite{QuantumTeleportation, MaNature2012, TakamotoNature2005, KesslerPRL2014, QuantumComputation, Chiaverini, Kelly}. 
A key point will be our ability to manipulate and control the fragile quantum systems, which requires high-fidelity fault-tolerant controls~\cite{Chiaverini, Kelly, ShorCode}. Quantum error correction, for example, needs computational infidelity below one part per ten thousand~\cite{fault tolerant thresholds}. 
It is necessary to develop robust quantum control methods that tolerate fluctuations coming not only from the environment but also from control parameters themselves.
When a two-level quantum system is controlled with coherent radiation, control errors are due to power fluctuation and frequency flickering. 

In recent years, various techniques, such as composite pulse sequences~\cite{composite pulse, sk composite pulse, chirp composite pulse}, pulse-shape programming~\cite{perturbation,shortcut}, and optimization techniques~\cite{grape, convex programming, Optimal Rabitz} have been proposed to achieve robust quantum controls, mainly addressing population inversion; these require either a train of well phase-maintained pulses or a complicated pulse shape, and/or reverse engineering.
Some of these techniques have been demonstrated in a microsecond radio-frequency regime when the pulses can be shaped directly in the time domain \cite{Mintert}. 

In this Letter, we demonstrate robust quantum control in the ultrafast femtosecond time-scale regime, when the shaping is operated in the frequency domain. In such systems, the frequency is relatively well stabilized ~\cite{Diddams}, therefore power fluctuation is the main source of error. The control scheme is designed to target a robust coherent superposition, i.e., with both the amplitude and the relative phase of the superposition made robust with respect to the power fluctuation. 
Quantum system driven by femtosecond chirped pulse is known to produce state selectivity~\cite{chirp RAP select,SongPRA2016} and robust population inversion~\cite{Chirp RAP theory, chirp RAP theory 2, chirp RAP QD} by rapid adiabatic passage ~\cite{RAP rev}. But there are very limited studies about robust creation of superposition of states. One can cite for instance the half Stark chirped rapid adiabatic passage technique \cite{lifting} in a nanosecond regime, which requires large pulse areas, and \cite{Mintert} in a microsecond radio-frequency regime.

We demonstrate the production of a robust superposition of arbitrary amplitude by deriving a simple and practical shaping involving only frequency quadratic chirping and static detuning from a single Gaussian pulse.  With a numerical investigation of the Sch\"{o}dinger equation (TDSE) and, as a proof-of-principle demonstration, a femtosecond laser-atom interaction experiment, we show the occurrence of a cusp in the final dynamics, which validates the existence of a robust control condition.
The physics behind this robust control may be understood in the context of a dynamical balance between the competing effects of chirping and detuning, where the former induces an adiabatic inversion and the latter its attenuation. 
 
The problem under consideration is the evolution of a two-level system driven by a chirped and detuned Gaussian pulse. The electric-field of the given pulse is defined in the frequency domain as 
\begin{equation}
E(\omega)=E_0\exp\left[{-\frac{(\omega-\omega_c)^2}{\Delta\omega^2}+i\frac{c_2}{2}(\omega-\omega_c)^2}\right],
\label{Efield}
\end{equation}
where $\omega_c$ is the center frequency of the pulse, $\Delta\omega$ the frequency bandwidth, and $c_2$ the frequency-domain chirp rate. The corresponding time-domain electric-field of this pulse features a Gaussian envelope with a linear chirp:
$E(t) = {\mathcal{E}(t)} \exp\left[{-i(\omega_c t +\alpha t^2 + \phi) }\right]/2 + c.c.$.

For a two-level system $\{|0\rangle,|1\rangle\}$ (of energies 0 and $\hbar \omega_0$, respectively), the Hamiltonian reads after the rotating frame transformation and the rotating wave approximation:
\begin{equation}
H(t) = \frac{\hbar}{2} \begin{pmatrix} 0 & \Omega(t) e^{-i \int \Delta(t) dt+i\phi} \\ \Omega(t) e^{i\int \Delta(t) dt-i\phi} & 0 \end{pmatrix},
\label{H0}
\end{equation}
where $\Delta(t)=\delta-2\alpha t$ is the instantaneous detuning with the static detuning $\delta=\omega_0-\omega_c$ and $\Omega(t)=\mu \mathcal{E}(t)$/$\hbar$ is the Rabi frequency with transition dipole moment $\mu$. 

Robustness of a dynamics, typically considered at the end of the pulse, is characterized by the second-order derivative, or the negative curvature, of the fidelity with respect to the considered fluctuation. Such quantity has been used for characterizing the composite sequences \cite{composite pulse, sk composite pulse, chirp composite pulse} or for single-shot robust pulses \cite{perturbation}. Robustness corresponds to a flat profile of the dynamics at the end of the pulse as a function of the fluctuation, i.e., to a very small absolute value of the curvature. Technically, it can be defined by 
a quantum geometric tensor~\cite{geometry ori, geometry}.
In our case, we consider the power fluctuation via the dimensionless error $\gamma=\delta \Omega /\Omega$ in Rabi frequency. 
The fidelity is defined by
$\mathcal{F}=|\langle \psi(\Omega) | \psi(\Omega+\gamma\Omega) \rangle |$ and the curvature $g$ is given by
\begin{equation}
g= -\frac{\partial^2\mathcal{F(\gamma)}}{\partial \gamma^2}\Big|_{\gamma=0}.
\label{QGT} 
\end{equation}
For a fluctuating Hamiltonian $H'=(1+\gamma)H$, the state vector initially at $|\psi({t=-\infty})\rangle = |0\rangle$ 
evolves pertubatively as
$U(t) \simeq U_{0}(t)-\frac{i}{\hbar}U_{0}(t)\int_{-\infty}^{t}\gamma V(t')dt'$,
where $U_{0}$ is the time-evolution matrix for $H$ and $V(t) = U_{0}^{\dagger} HU_{0}$. 
We obtain
\begin{equation}
g = \langle 0| \left(-\frac{i}{\hbar}\int_{-\infty}^{\infty}Vdt \right)^{\dagger} | 1 \rangle \langle 1 | \left(-\frac{i}{\hbar }\int_{-\infty}^{\infty} V dt \right) | 0\rangle   
\label{g_perturb}
\end{equation} 
up to the second order of the fluctuation $\gamma$ \cite{perturbation}.

\begin{figure}[thb]
\centerline{\includegraphics[width=0.45\textwidth]{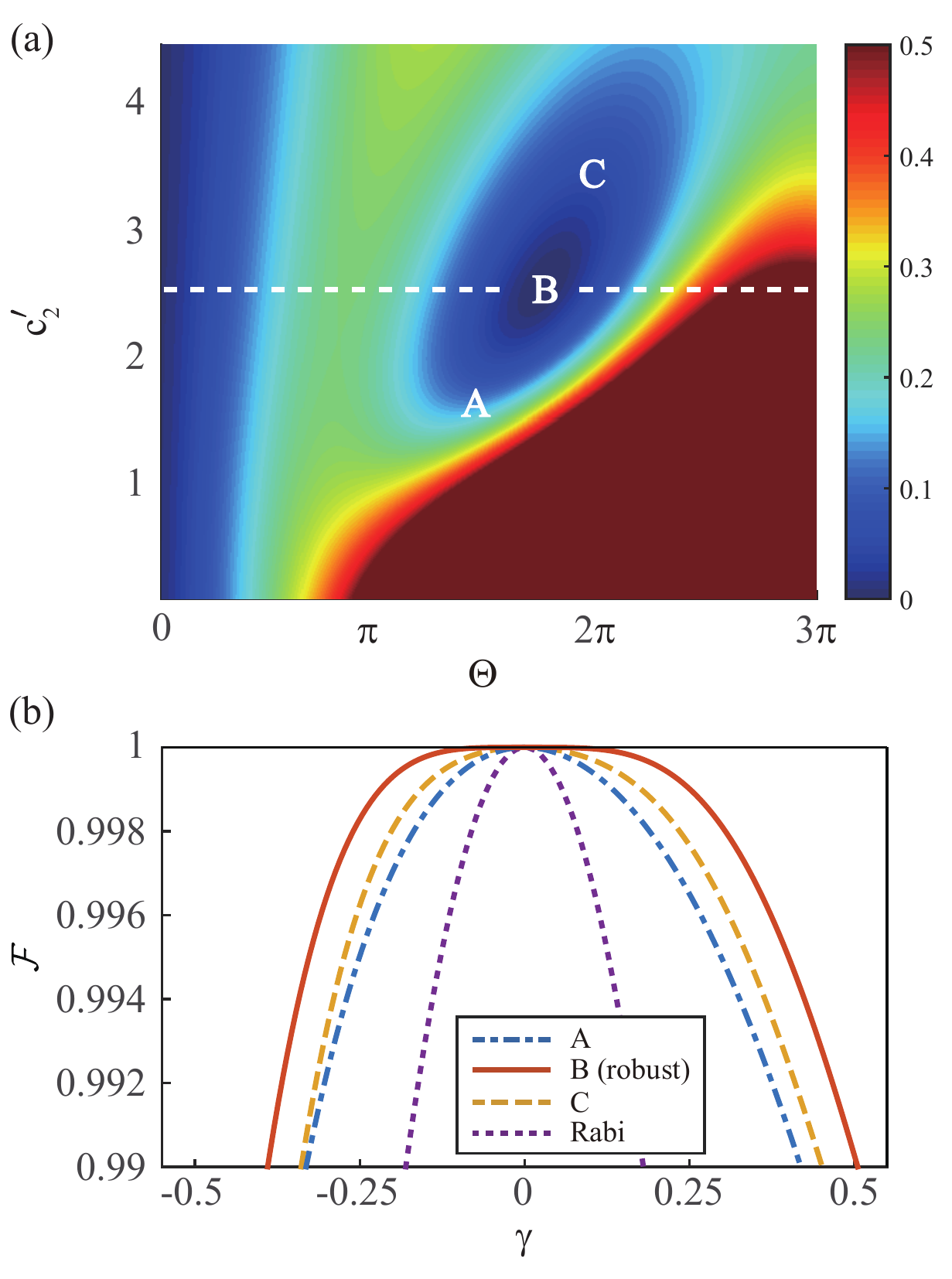}}
\caption{(Color online) Robustness map $g(\Theta, c_2')$ for ${\Delta'=0.637}$. The negative curvature $g$ is plotted as a function of the pulse area $\Theta$ and the chirp rate $c_{2}'$. The most robust point is located at \textbf{B}, while $g_{\textbf A}=0.11$, $g_{\textbf C}=0.05$, and $g_{\rm Rabi}=0.62$.
 (b) Fidelity curves $\mathcal{F}(\gamma)$: The fidelity curve (red solid line) is calculated at \textbf{B} as a function of the fluctuation $\gamma$ (see text for definition) and compared with cases  \textbf{A} and \textbf{C} (blue dash-dotted and yellow dashed lines, respectively) and also with the Rabi-type evolution (purple dotted line) for the same excited-state probability $P_e=0.5$.}
\label{fig1}
\end{figure}

Figure~\ref{fig1} shows the numerical calculation of the robustness using the curvature $g$ and the fidelity $\mathcal{F}$ for the two-level system dynamics driven by detuned and chirped pulses. To make the comparison easier, we use the dimensionless parameters $\Delta'=\delta/\Delta\omega$, $c_2'=c_2 \Delta\omega^2$, and the pulse-area (after shaping) $\Theta=\int^{\infty}_{-\infty}\Omega(t) dt$. 
The curvature $g(\Theta, c_2')$, is plotted for a particular value $\Delta'=0.637$ in Fig.~\ref{fig1}(a), which has been chosen such that the dynamics reaches at the end of the pulse a coherent superposition with equal weights $P_e=0.5$. Optimal robustness ($g\approx 0$) occurs at point \textbf{B}: $(\Theta, c_2', \Delta')=(1.78\pi,  2.52, 0.637)$. Less robust pulses with a smaller or larger chirp rates are also shown in Fig.~\ref{fig1}(a), respectively denoted by \textbf{A} and \textbf{C}, which have small but not nonzero curvatures. 
We remark that $g=0$ points for smaller or larger chirp rates may also be found at different $\Delta'$s and $\Theta'$s, but occurring for larger pulse areas.
Therefore, as shown in Fig.~\ref{fig1}(b), the optimal robust pulse \textbf{B} exhibits a significantly flattened fidelity curve, or more robust evolution, than other pulses and the Rabi-type evolution. The optimal pulse \textbf{B} shows a 3.5 times wider range of robust evolution than the Rabi oscillation. 

\begin{figure}[thb]
\centerline{\includegraphics[width=0.4\textwidth]{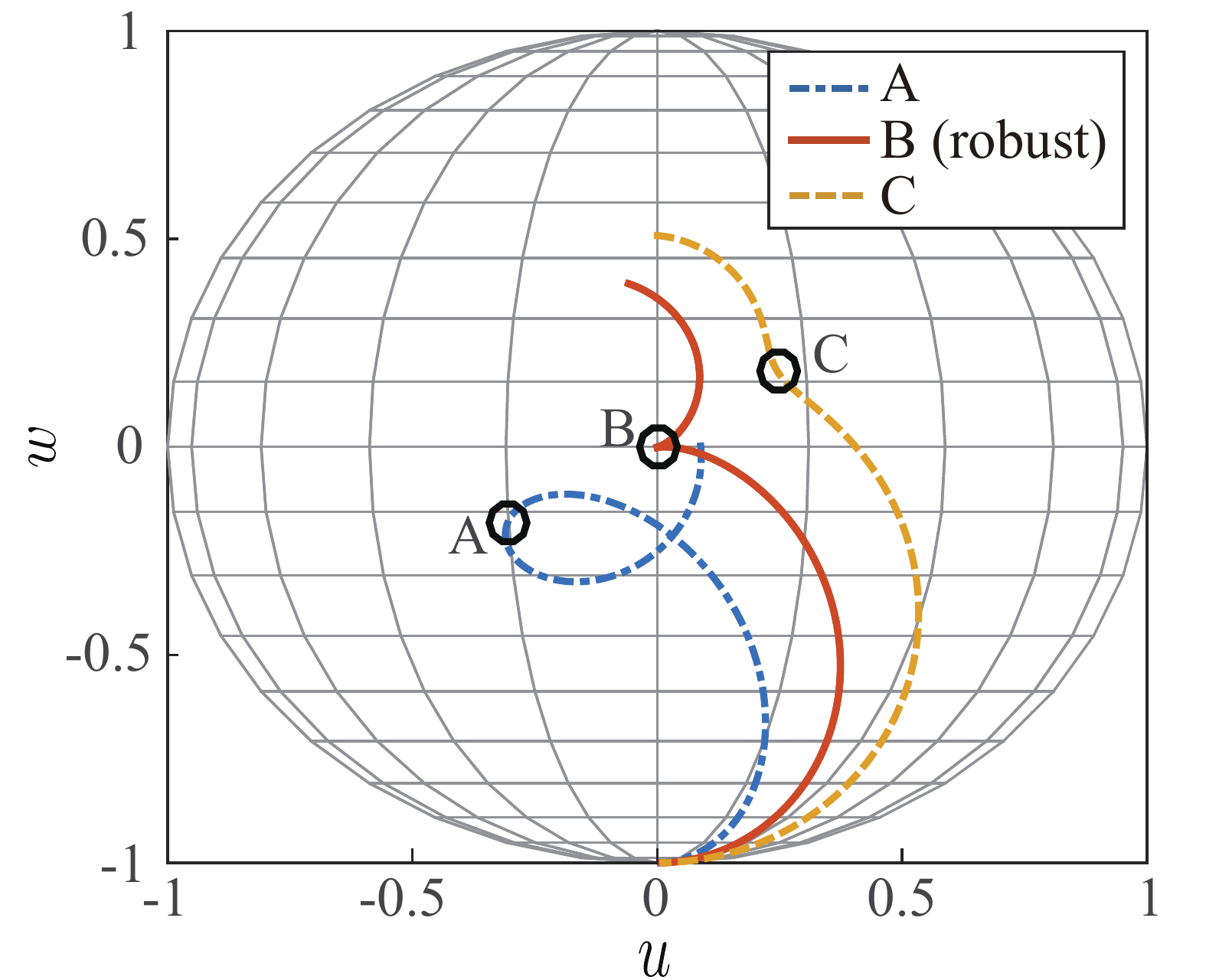}}
\caption{(Color online) $\Theta$-parameterized trajectories on Bloch sphere: The final state after a chirped- and detuned-pulse interaction is plotted with a Bloch vector. The trajectories through \textbf{A}, \textbf{B}, and \textbf{C} in Fig.~\ref{fig1} are plotted with blue dash-dotted, red solid, yellow dashed lines, respectively.}
\label{fig2}
\end{figure}

The horizontal line segment in Fig.~\ref{fig1} corresponds to a $\Theta$-trajectory on the Bloch sphere. 
Figure~\ref{fig2} shows such trajectories passing through respectively \textbf{A}, \textbf{B}, and \textbf{C} points, which are plotted as a function of $\Theta$, while $c_2'$ and $\Delta'$ are fixed in each trajectory. As clearly shown in Fig.~\ref{fig2}(a), the trajectory shape changes, as $c_2'$ increases, from a looped curve (case \textbf{A}) to an unlooped one (case \textbf{C}), and, as a result, a cusp is formed in-between (case \textbf{B}). The trajectory with a cusp is particularly interesting in topology, because the singular nature of the cusp allows both derivatives of any pair of mutually-orthogonal coordinates, on the Bloch sphere, with respect to $\Theta$ being always zero~\cite{cusp}, i.e., $\frac{d|\psi\rangle}{d\Theta}=\frac{\partial |\psi\rangle}{\partial \theta} \frac{d\theta} { d\Theta} +\frac{\partial |\psi\rangle}{\partial \phi} \frac{d\phi}{d\Theta} =0$ at the cusp point. This occurrence ensures the curvature to be zero (the optimal robustness). For higher pulse areas (and different chirp rates), other similar cusps occur associated to other robust points (not shown). 
 

To demonstrate the robustness experimentally, we performed femtosecond laser-atom interaction experiments. The experimental set-up and procedure are similar to those in our previous experimental work~\cite{LimSR2014, LeeOL2015, LeePRA2016, SongPRA2016}.  In brief, a magneto-optical trap (MOT) was used to confine rubidium atoms ($^{85}$Rb) in a small volume for uniform laser interaction. The diameter of the atomic vapor inside the MOT was 300~$\mu$m, about 43\% of the laser diameter. The laser setup consisted of a femtosecond laser amplifier and, as a pulse-shaping device, an acousto-optic programmable dispersive filter (AOPDF)~\cite{AOPDF}. Femtosecond laser pulses were initially produced from  a mode-locked titanium-sapphire laser oscillator and amplified up to 0.85~mJ of single-pulse energy at a repetition rate of 1~kHz.  Each laser pulse was then shaped with four experimental parameters: center frequency, chirp rate, bandwidth, and pulse intensity. The first three were programmed with the AOPDF and the last, the laser intensity, was fine-controlled with a half-wave plate sandwiched between a pair of cross-polarizers. The center wavelength of the laser pulse was tunable from $\lambda_c = 792$ to 802~nm, which corresponded to the detuning range between $\delta = -8.38\times 10^{12}$ and $2.13\times 10^{13}$~rad/s. The laser bandwidth was fixed at $\Delta \lambda_{\rm FWHM}=10.4$~nm ($\Delta\omega_{\rm FWHM} = {3.1\times 10}^{13}$~rad/s),
and the frequency chirp rate $c_2$ was changed from -40,000 to 40,000~fs$^2$ for various experiments. 
The two-level system was formed with 5S$_{1/2}$ and 5P$_{1/2}$, the ground and the first-excited states of atomic rubidium ($^{85}$Rb). The population leakage to other states, including 5P$_{3/2}$, 5D, and ionization levels, was less than 2\%
within the experimental parameter range. After the atoms were controlled by the as-shaped laser pulse, those in the excited state were ionized by a probe laser pulse, which was the frequency-doubled split-off from the unshaped laser pulse, and measured with a micro-channel plate detector. The total sequence of experiment was tuned at 2~Hz cycle to maintain the MOT density by using mechanical shutters for femtosecond laser pulses, and acousto-optic modulators for MOT lasers. The MOT lasers were turned off before the arrival of the control pulse to initialize the atomic state in the ground state and turned on after the interaction to restore the MOT. 

\begin{widetext}

\begin{figure}[thb]
\centerline{\includegraphics[width=1.0\textwidth]{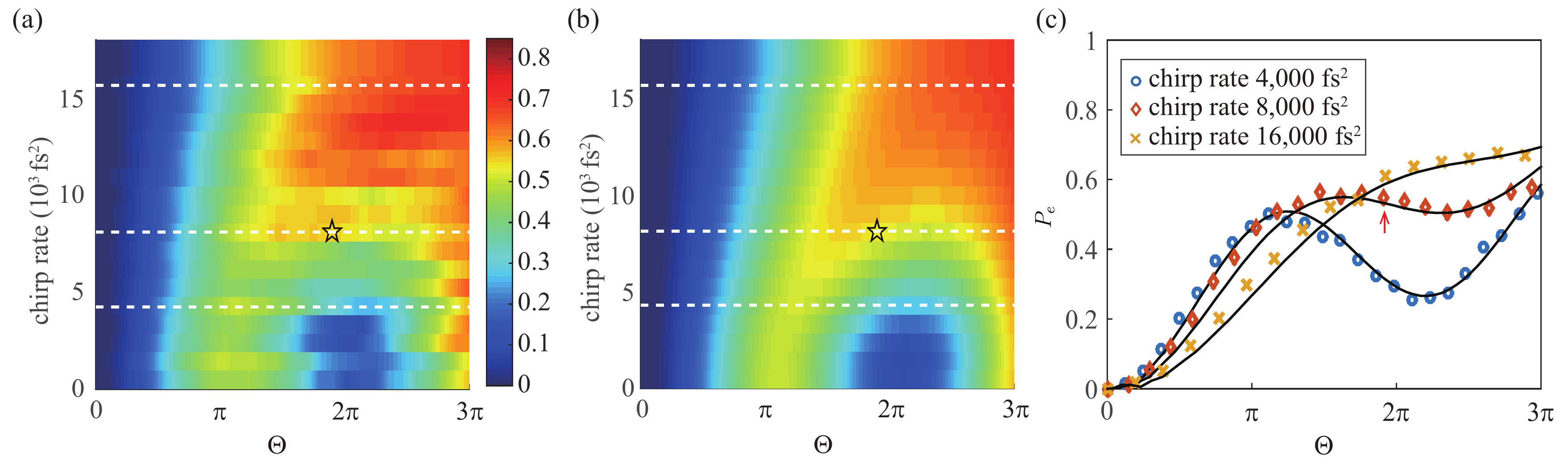}}
\caption{(Color online) The probability $P_e(\Theta, c_2)$ of the excited-state (5P$_{1/2}$ of rubidium) after the detuned and chirped Gaussian pulse excitation as a function of the pulse-area $\Theta$ and the chirp rate $c_2$, while the detuning and laser bandwidth are fixed at $\delta=3.5$~nm and $\Delta\omega_{\rm FWHM}= 3.1\times 10^{13}$~rad/s, respectively.  (a) Experimentally obtained probability map $P^{\rm exp}_e(\Theta,c_2)$, to which the measured atom population was converted using Rabi-oscillation calibration measurements (see text for detail). (b) Theoretical result  $P_e^{\rm TDSE}(\Theta, c_2)$, obtained using the TDSE calculation for an atom cloud of a Gaussian profile with a diameter 47\% of the Gaussian laser beam diameter. (c) The behavior of the population $P_e(\Theta)$ for selected chirp rates $c_2=4,000$ (blue circles), $8,000$ (red diamonds), and $16,000$~fs$^2$ (yellow crosses), respectively shows good agreement with the TDSE calculation (solid lines). The star and arrow symbols represent the robust condition.} \label{fig3}
\end{figure}

\end{widetext}

Experimental results are compared with numerical calculation in Fig.~\ref{fig3}.
The excitation probability $P_e(\Theta, c_2)$ of atoms after the shaped laser pulses was probed as a function of the pulse-area $\Theta$ and the chirp rate $c_2$, while the detuning was fixed at $\delta=3.5$~nm ($\Delta'=0.56$). This detuning corresponds to a coherent superposition with $P_e=0.6$. The results are shown in Fig.~\ref{fig3}(a). We note that, to retrieve $P_e^{\rm exp}(\Theta, c_2)$ from the measured counts of the ionized electrons, we used Rabi-oscillation calibration method~\cite{LeeOL2015}. The ideal robust control point, marked with star in the figures is located at $(\Theta, c_2', \Delta')=(1.9\pi,2.79,0.56)$ or $(\Theta, c_2, \Delta)=(1.9\pi,8.1\times 10^{3}$~fs$^2,  1.04\times 10^{13}$~rad/s).
In addition, we assumed the minor discrepancy at the high laser-power region { ($\Theta>2.5\pi$)} was attributed to the effect of a possible pre-pulse with 0.4\% energy and small relative phase. The result of TDSE simulation for the two-state system dynamics is shown in Fig.~\ref{fig3}(b), where the spatial inhomogeneity~\cite{LeeOL2015} of the laser-atom interaction is taken into account. 
Figure~\ref{fig3}(c) shows that the overall behavior of the two-level system dynamics is in good agreement with the experimental data.

\begin{figure}[thb]
\centerline{\includegraphics[width=0.48\textwidth]{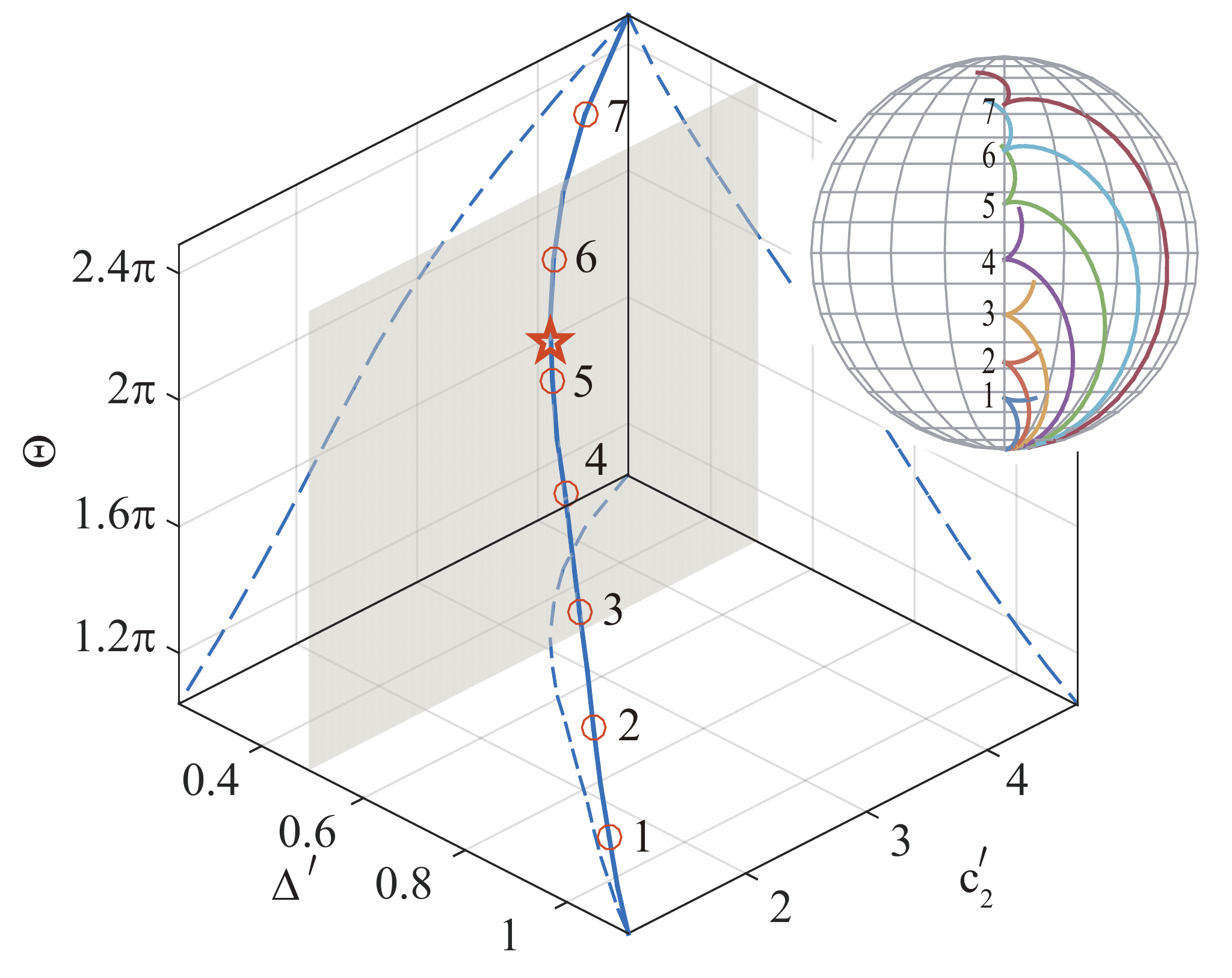}}
\caption{(Color online) Robust control conditions in the parameter space ($\Delta'$, $c_2'$, $\Theta$): Red circles represent the parameters for detuned and chirped pulses that induce robust control to various target probabilities; the solid and dashed lines are the numerical fit and its projections to each plane, respectively. The inset shows the $\Theta$-trajectories on the Bloch sphere, corresponding to the robust control red circles, each featuring a cusp.}
\label{fig4}
\end{figure}

We now turn our attention to the generalization of our robust control method to arbitrary target states. We experimentally probed the two-dimensional section at $\Delta'= 0.56$ of the three-dimensional parameter space of ($\Delta'$, $c_2'$, $\Theta$) as illustrated in Fig.~\ref{fig4}, where the targeted robust point is marked with a star. Further numerical investigations show that robust control conditions occur along a line (the solid line with circles) in the parameter space, where each value corresponds a specific value of $P_e$. 
Along the line, the control parameters are numerically fitted to a function of $P_e$ in the range from 0.08 to 0.98 as given in TABLE.~\ref{Table1}. The result indicates that the robust control can be made to arbitrary target probabilities. By considering the fact that the azimuth angle of the Bloch vector, or the relative phase of the superposition, can be easily set with the carrier-envelope phase of the laser pulse~\cite{LimSR2014}, our method can be thus generalized to any target state, of arbitrary amplitude and phase, on the Bloch sphere. 

\begin{table}[thb]
\caption{Robust control conditions numerically fitted to $P_e= A+B/({1+C e^{-D x}})$, a step-like function of $P_e$, where $x=\Delta'$, $c_2'$, or $\Theta/\pi$.}
\begin{ruledtabular}
\begin{tabular}[c]{c c c c c}
x & $A$ & $B$ & $C$ & $D$\\
\hline
$\Delta'$ & -0.055 & 1.19  & 0.079 & -4.20 \\
$c_2'$ & -0.097 & 1.076 & 22.5 & 1.32 \\
$\Theta/\pi$ & $-0.0033$ & 1.019 & 264 & 3.14\\
\end{tabular}
\end{ruledtabular}
\label{Table1}
\end{table}

Our strategy produces robust coherent superpositions, far from the inversion, such as the typical half coherent superposition, and it can be achieved in an ultra-fast pulse duration. The large pulse area limit in Fig.~\ref{fig4} shows that the population inversion is recovered in the adiabatic limit (however without modification of the energy of the initial pulse). For more complicated systems, we expect more parameters for the shaping to achieve robustness, which will result in a non-linear chirp and non-Gaussian pulse in general (see for instance \cite{stark1, stark2, stark3, shaping,shaping2}). 
If other states come into play and perturb our two-level system, the strategy will consist in treating them by adiabatic elimination. This will result in a dynamical Stark shift, corresponding to an additional detuning that can be incorporated in the Hamiltonian and compensated by the chirping. Further work will consist of adapting the shaping in order to produce a robust qubit gate such as the Hadamard gate.


In summary, we have shown that detuned and chirped pulses can implement robust transfer of a ground-state to a chosen coherent superposition of arbitrary amplitude and phase. Our numerical study proves that robustness is associated to a cusp in the final dynamics as a function of the power fluctuation. As a proof-of-principle experiment, we performed ultrafast optical control of cold rubidium atoms, which validates the numerical simulations. Our robust control solution is particularly simple, requiring only frequency quadratic chirping (i.e., temporal linear chirping) and static detuning from a Gaussian-shape pulse. Producing such robust coherent superpositions at the femtosecond timescale is anticipated to become useful in a wide range of applications.

\begin{acknowledgements}
The authors are grateful to C. H. Raymond Ooi, Yunheung Song, and Adam Massey for fruitful discussions.
This research was supported by Samsung Science and Technology Foundation [SSTF-BA1301-12]. 
\end{acknowledgements}

\end{document}